\documentclass{elsart}

\usepackage{natbib}

 \usepackage{graphics}

\begin{document}

\begin{frontmatter}



\title{The turbulent interstellar medium}


\author{A. Burkert}

\address{University Observatory Munich, Scheinerstr. 1, D-81679 Munich, Germany}
\ead{burkert@usm.uni-muenchen.de}

\begin{abstract}
An overview is presented of the main properties of the interstellar medium.
Evidence is summarized that the interstellar medium is highly turbulent,
driven on different length scales by various energetic processes.
Large-scale turbulence determines the
formation of structures like filaments and shells in the diffuse interstellar medium.
It also regulates the formation of dense, cold molecular clouds.
Molecular clouds are now believed to be transient objects that form on
timescales of order $10^7$ yrs in regions where HI gas is compressed and cools. 
Supersonic turbulence in the compressed HI slab is generated by a combination
of hydrodynamical instabilities, coupled with cooling. Turbulent dissipation is
compensated by the kinetic energy input of the inflow. Molecular hydrogen eventually
forms when the surface density in the slab reaches a threshold value of
$\sim 10^{21}$ cm$^{-2}$ at which point further cooling triggers the onset of
star formation by gravitational collapse. A few Myrs later, the newly formed
stars and resulting supernovae will disperse their molecular surrounding and generate new expanding
shells that drive again turbulence in the diffuse gas and trigger the formation
of a next generation of cold clouds.
Although a consistent scenario of interstellar medium dynamics and star formation
is emerging many details are still unclear and require more detailed work on 
microphysical processes as well as a better understanding 
of supersonic, compressible turbulence.

\end{abstract}


\end{frontmatter}

\section{Introduction}
\label{introduction}
Steady-state multi-phase models have dominated our picture of the interstellar
medium (ISM) in the Milky Way for a long time. According to the early 
model of Field et al. (1969) and subsequent modifications 
the ISM represents an ensemble 
of two stable gas phases that are in thermal pressure equilibrium
with a mean pressure of n $\times$ T $\approx$ 1000 K cm$^{-3}$.
Cold molecular clouds with mean densities of
n $\approx$ 100 cm$^{-3}$ and temperatures T $\approx$ 10 K are embedded in a warm, diffuse
and partly ionized intercloud component with density n $\approx$ 0.1 cm$^{-3}$ and temperatures of 
order $10^4$ K. The gas clouds contain a large fraction of the total mass and move as stable,
dense spheroidal objects in the widespread intercloud medium which on the other hand has 
the dominant volume filling factor. Star formation would eventually heat
and disperse the massive clouds. New generations of small clouds form from the 
intercloud medium by cooling instabilities (Field 1965) and subsequently grow
by random inelastic cloud-cloud collisions (e.g. Elmegreen 1989).
Already in 1977 McKee \& Ostriker noticed that the two-phase model
could not be valid as supernova explosions should lead to a third,
hot and tenuous gas phase. However they still focussed on the importance of thermal
pressure equilibrium and a steady state description as the main physical constraint
to evaluate the state of the various gas phases.

This situation has changed drastically in the last decade. High-resolution observations
e.g. with the Infrared Astronomical Satellite (IRAS) and more recently with the
Spitzer satellite reveal a complex kinematical state and spatial structure of 
the interstellar gas. The ISM appears to be far from hydrostatic
equilibrium and turbulent. In fact, it is the kinetic turbulent
pressure, not thermal pressure, that probably dominates the gas dynamics and that
characteristizes its density structure and the dynamics of the various gas phases.
Turbulence couples structures on very different scales. Molecular clouds might just
represent the high-density tail of this hierarchy, forming
in colliding gas flows that lead to transient local compressions that subsequently cool.
We will discuss below that the complex internal structure of molecular clouds 
that determines their evolution and their condensation into stars and stellar clusters
is also a result of turbulence, generated by various hydrodynamical instabilities 
during the process of molecular cloud formation.

With turbulence becoming the dominant source of structure in the ISM, theoretical models 
lost their simplicity and equilibrium descriptions had to be replaced by dynamical models 
where large and small scales are simultaneously considered. We are just starting to explore and 
understand this rich and enormously complex new 
field of astrophysics.  In this short review I can only focus on a few
interesting topics and unsolved questions. 
An excellent and comprehensive review of our current understanding
is presented by Elmegreen \& Scalo (2004) and Scalo \& Elmegreen (2004). 
Reviews that focus especially on numerical simulations of ISM turbulence
are e.g. V\'{a}zquez-Semadeni et al. (2000) and Ballesteros-Paredes et al. (2006). 
A summary of our understanding
of star formation in turbulent clouds is given by Mac Low \& Klessen (2004).

\section{Turbulence in the diffuse ISM}
\label{driver}

\begin{figure}
\resizebox{12cm}{!}{
{\includegraphics{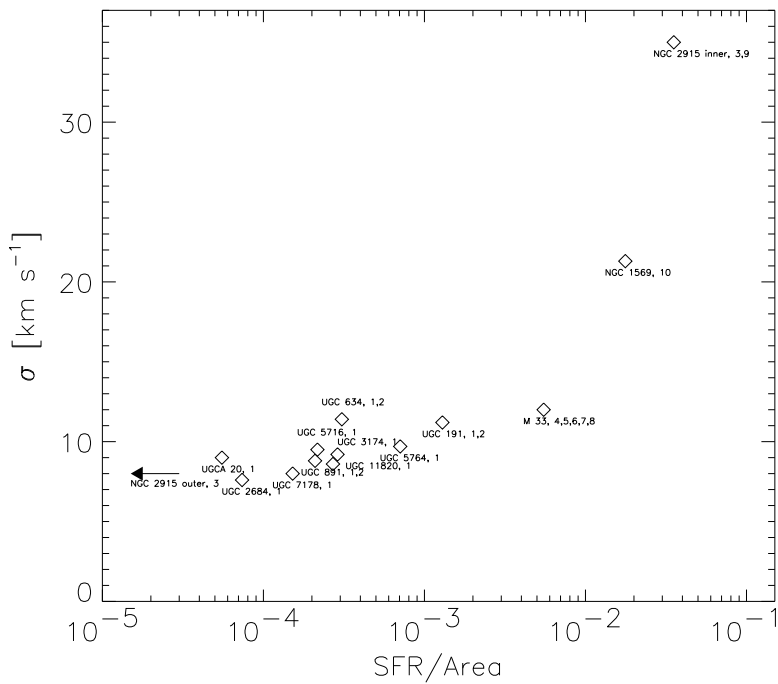}} 
}
\caption{Characteristic velocity dispersion $\sigma$ of a sample of disk galaxies
as function of their average surface star formation rate in units of (M$_{\odot}$  yr$^{-1}$
kpc$^{-2})$ (Dib et al. 2006). $\sigma$ is almost independent of the star formation rate
for typical values found in slowly evolving disk galaxies like the Milky Way. It rises
steeply in star burst regions with high star formation rates.}
\label{fig1}
\end{figure}

It is now a well established fact that the ISM in galactic disks is dominated by irregular and often supersonic
gas motions (Larson 1981; Scalo 1987; Dickey \& Lockman 1990). In most spiral galaxies HI emission lines exceed
the values, expected from thermal broadening, indicating turbulent velocity dispersions 
$\sigma$ of order 10 km/s. The velocity dispersion in galactic disks in general decreases
outwards from $\sigma \approx$ 12-15 km/s in the inner regions to $\sigma \approx$ 4-6 km/s in the
outer parts. Figure 1 shows the characteristic velocity dispersion for
a sample of galaxies as function of their surface averaged star formation rate (Dib et al. 2006).
Note that self-regulated star formation rates that are typical for Milky Way-type disk galaxies 
lead to velocity dispersions of order
$\sigma \approx$ 6-8 km s$^{-1}$, independent of the star formation rate.
The diffuse ISM acts like a thermostate. The situation
changes drastically in situations where star formation is getting out of control, leading to starbursts that last
a short time, of order $10^8$ yrs. As shown in figure 1, the transition into the starburst
regime is marked by a steep increase of the velocity dispersion.

\subsection{Driving turbulence in the diffuse ISM}
Several physical processes, acting on different scales and injecting different amounts of
kinetic energy contribute to the driving of ISM turbulence. However, despite a large amount of
numerical work in this field, the dominant energetic sources
and the physical processes that convert the kinetic energy into turbulence are not well understood.
Stars are obvious candidates. Large-scale
expanding gas flows could e.g. be generated by high-pressure HII regions, resulting from the UV
radiation of young, massive stars (Kessel-Deynet \& Burkert 2003), stellar winds or supernova explosions. 
Mac Low \& Klessen (2004) argue that supernova explosions dominate the global kinetic energy input
into the interstellar medium. Several two- and three-dimensional numerical simulations 
have tested supernova driving in galactic disks 
(e.g. Kim et al. 2001; de Avillez \& Breitschwerdt 2004, 2005; Slyz et al. 2005; Mac Low et al. 2005).
The recent investigation by Dib et al. (2006) demonstrates that the HI gas velocity dispersion 
saturates at values of 3 km/s for values of the supernova rate, ranging from 0.01 to
0.5 the Galactic value (1/57 yr$^{-1}$, Cappellaro et al. 1999). It increases
sharply at larger rates, reproducing the transition into the starburst regime. Although the constant
velocity dispersion, found at lower rates is promising, the actual 
value of 3 km/s is a factor of 2-3 lower than observed. Part of this discrepancy
might be due to thermal line broadening. However other feedback processes probably
are also important to produce the observed level of turbulence in galactic disks.

What is the origin of a constant HI velocity dispersion, independent of the supernova rate?
Dib et al. (2006) argue that the answer lies in a dominant gas phase
that is in a thermally unstable temperature regime
between 400 K $\leq$ T $\leq$  10000 K. It is produced when individual supernova remnants cool
and are dispersed. In these regions thermal instability (Burkert \& Lin 2000) generates 
local irregular pressure gradients that lead to HI gas flows with typical velocities of order
the local sound speed, corresponding to a few km/s. The importance of thermal instability as
a driver of turbulence in the ISM has previously been discussed in details by Kritsuk \& Norman
(2002a,b). In the starburst regime, on the other hand, supernova remnants begin to overlap
in an early hot phase, generating a stable, hot gas component with a large volume filling factor.
HI flows are now directly coupled to the expansion of bubbles generated by multiple supernovae,
leading to larger velocities that are however still subsonic with respect to the hot gas component.

Note, that in the Milky Way this scenario would predict the existence of a 
thermally unstable gas phase with a high mass fraction
(Gazol et al. 2001), in contrast to previous static multi-phase models. 
A large fraction of interstellar gas in the unstable regime
has indeed been observed (Dickey et al. 1977; Heiles 2001; Kanekar et al. 2003).
As outlined by Elmegreen \& Scalo (2004), this phase also explains the origin of
large variations in observed gas pressures that were puzzling in the static models
(Jenkins 2004; Kim et al. 2001).

Many physical mechanisms, not related to stellar energetic feedback, could in principle 
contribute to the driving of turbulence in the ISM. Clear evidence for additional sources
is for example the high HI velocity dispersion observed in the outer parts of
galactic disks where star formation is negligible (Dickey et al. 1990).
Numerical simulations are just starting to explore these drivers of turbulence
in greater details.
Galactic rotation, for example, represents a huge reservoir of kinetic energy. Wada et al. (2002)
demonstrate that the dissipation of turbulent energy in disks could be compensated
by a combination of galactic shear and gas self-gravity. In addition, the coupling of
galactic shear with magnetic fields can trigger a magnetorotational instability (MRI; Balbus \& 
Hawley 1991; Sellwood \& Balbus 1999). Three-dimensional simulations by Kim et al. (2003) show
that the MRI could generate velocity dispersions of order
a few km/s which is similar to supernova driving (see also Dziourkevitch et al. 2004). 
Similar values are found by Piontek \& Ostriker (2004) who studied the combined affect of
thermal instability and MRI.

Only recently have galaxies in early phases of evolution been detected in
deep images with the Hubble Space Telescope (e.g. Cowie et al. 1995;
Tran et al. 2003; Elmegreen et al. 2004, Forster Schreiber et al. 2006). They are characterised by a few
giant blue, bright clumps of sizes $\sim$ 500 pc where stars appear to form with
high efficiency in a starburst mode. The clumps dominate the disk light, suggesting
a global gravitational disk instability and implying an
unusually high velocity dispersion of order 30\% the disk rotational velocity
(Elmegreen 2004), corresponding to
40-60 km/s which is similar to the velocity dispersion of the thick disk component
of the Milky Way. The triggering mechanism of this highly turbulent starburst mode in young galaxies
and its affect on the evolution of the various galactic components and
galaxy morphologies is not well understood up to now.

\subsection{HI holes and the complex filamentary structure of the ISM}

\begin{figure}
\resizebox{18cm}{!}{
{\includegraphics{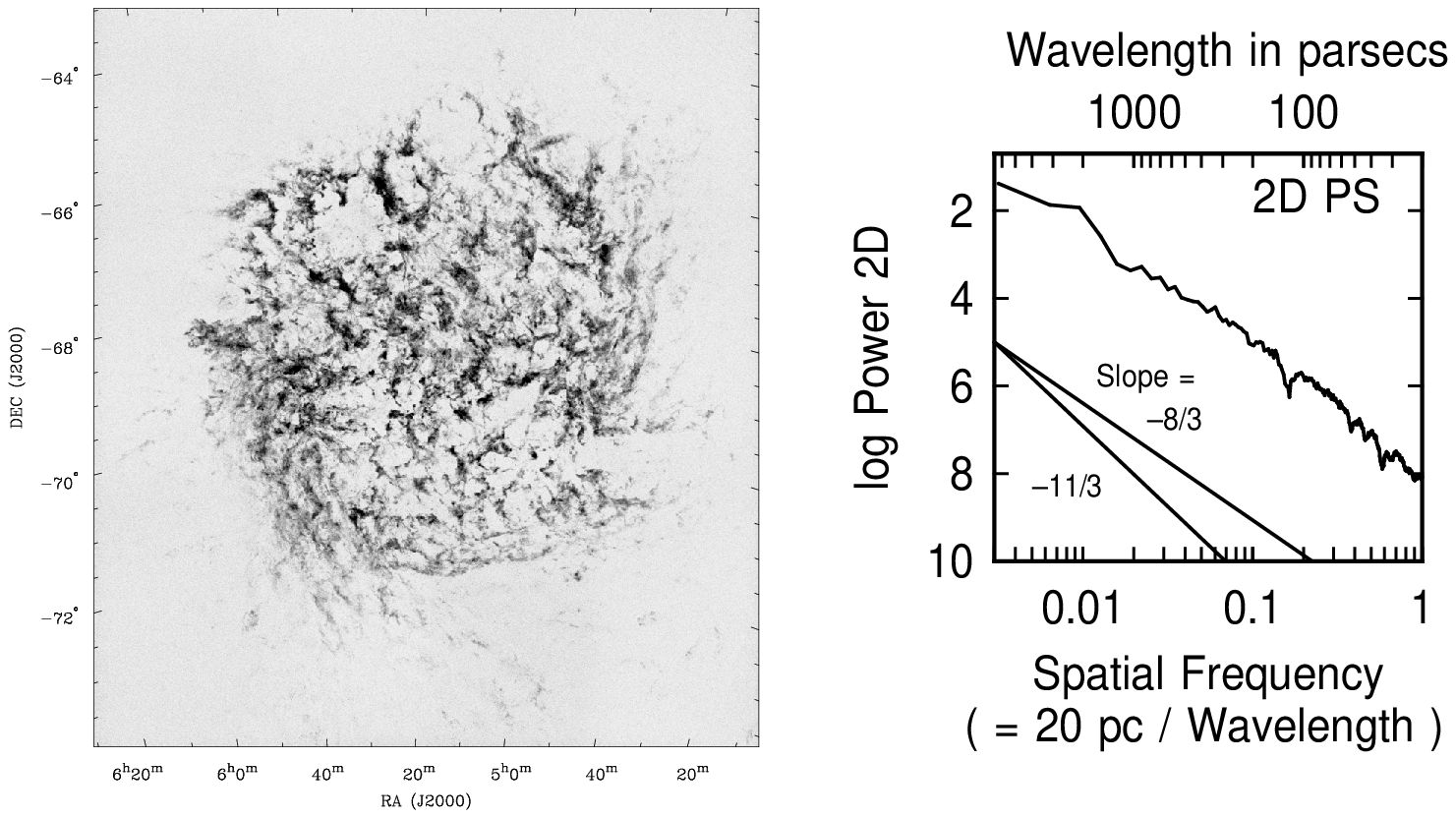}} 
}
\caption{The left figure, provided by S.Kim (Kim et al. 1999),
shows the peak 21 cm HI surface density distribution in the Large
Magellanic Cloud which nicely demonstrates the existence of a complex network
of filaments and shells. The right figure, provided by B. Elmegreen 
(Elmegreen et al. 2001), shows
the  power spectrum of the HI emission in the LMC which is well fitted by a
power-law of slope -3 over 2 decades in length.}
\label{fig2}
\end{figure}

The left panel of figure 2 shows the peak 21cm HI surface density distribution in the Large Magellanic Cloud
(Kim et al. 1999) which reveals a complex network of interacting filaments, shells and superbubbles.
The question of which physical processes produce these holes in the LMC and other galaxies is not solved.
Their circular shapes suggest a stellar central energy source. 
Observations however indicate that the voids especially in the outer parts of galactic disks
are often not a result of supernova explosions. Kim et al. (1999) for example find only a weak
correlation between the positions of the HI shells in the LMC and HII regions. Rhode et al. (1999)
have studied 51 HI holes in the dwarf galaxy Holmberg II (Puche et al. 1992). They show that in 
86\% of all cases the holes do not show any signature of the presence of an embedded stellar cluster
or any sign of ongoing stellar activity. In fact, X-ray observations of Holmberg II show that
the HI holes are often devoid of hot gas (Kerp et al. 2002) and therefore probably did not form
by the expansion of a hot gas bubble, sweeping up its environment.

Dib \& Burkert (2005) suggest that large HI holes can form naturally as a combined result of
ISM turbulence, coupled with thermal and gravitational instabilities. Their hydrodynamical simulations
of large-scale driven turbulence, including cooling and heating processes as well as self-gravity
can reproduce the structure of shells and holes, observed in regions where no stellar activity is
observed. For a more quantitative analyses they subdivided the gas disk into rectangular cells of constant
size $l$ and determined the autocorrelation lengthscale of
the HI surface density distribution in each cell. Averaging over all cells, the
mean autocorrelation length scale 
$l_{cr}$ was determined as function of cell size $l$. Dib \& Burkert find that $l_{cr}$
increases linearly with $l$ as long as the cell size is smaller than the length scale
$l_{turb}$ on which
turbulent energy is injected into the ISM. Once $l > l_{turb}$,
the autocorrelation lengthscale becomes independent of the map size. This analysis can be
used to determine the scale of energy injection into the ISM for observed HI disks. 
Applying the method to Holmberg II leads to a driving scale of $l_{turb} \approx$ 6 kpc which is 
very puzzling as this scale is much larger than any known energy source in the galaxy.

The power spectra of 2D gas column densities and emission fluctuations are often fitted well
by power-law profiles with a slope around -3 (right panel of Fig. 2, see Elmegreen et al.
2001). This power-law extends from the largest to the smallest observable scales which is surprising
given the fact that multiple sources of energy are likely to contribute to the driving of turbulence on
very different length scales. The similarity of this slope to a 2D Kolmogorov power spectrum
(Kolmogorov 1941) of -8/3 is also not well understood as Kolomogorov's scaling relations are 
striktly valid only for incompressible fluids where vortices (solenoidal modes) are the relevant dynamical
structures in contrast to the ISM where compressible
modes leading to shocks and rarefaction waves are important. As the Fourier transform of a step function
has a 2D power-law slope of -3 (Mac Low \& Klessen 2004) we might just see the complex network of
interacting shock fronts. Unfortunately numerical simulations of the turbulent supersonic ISM do not
have enough dynamical range yet to investigate this interesting question in greater details.

\section{Turbulence in the dense, cold interstellar medium}

\begin{figure}
\begin{center}
\resizebox{10cm}{!}{
{\includegraphics{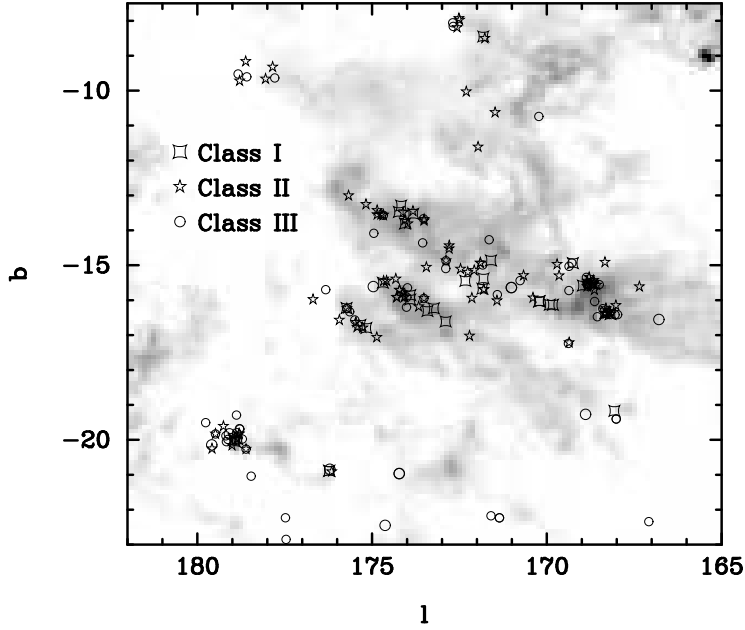}} 
}
\end{center}
\caption{The distribution of young stars in the Taurus molecular cloud,
superimposed upon $^{12}$CO emission. Most of the stars have ages of order 2 Myr
despite the fact that the lateral sound crossing time is of order 20 Myr 
(figure provided by L. Hartmann; see Hartmann et al. 2001 and Hartmann 2002, 2003).}
\label{fig3}
\end{figure}

Stars form in the cold, dense molecular phase of the interstellar medium where 
clumps with masses in the stellar regime can
become gravitationally unstable and collapse. Like the diffuse ISM, molecular clouds (MC)
exhibit a wealth of clumpy and filamentary substructures that indicate that they
are again turbulent regions, embedded and interacting with the turbulent diffuse interstellar
medium.

\subsection{The problem of star formation in a crossing time}

How stars form in MC is an important unsolved question of modern astrophysics
(for a review see Mac Low and Klessen 2004).
Ordinary spiral galaxies like the Milky Way form stars at a low, self-regulated rate. Although a
large fraction of the visible gas is condensed in MCs with masses in the range of 
$10^4 - 10^6$ M$_{\odot}$ that by far exceed their thermal Jeans mass, star formation turns out
to be surprisingly inefficient (Blitz \& Shu 1980). The Milky Way, for example, with a total
molecular gas mass of order $2 \times 10^9$ M$_{\odot}$ and mean molecular cloud densities
of order 100 cm$^{-3}$, corresponding to collapse timescales of $5 \times 10^6$ yrs, could
in principle form stars with a rate of more than 100 M$_{\odot}$/yr which is a factor of 100 larger than 
observed. 

Supersonic (Mach numbers: M $\approx$ 5-10) turbulent gas motions have been detected in 
most cloud complexes (Larson 1981, Falgarone \& Philips 1996; 
Elmegreen \& Falgarone 1996, Williams et al 2000) and are considered 
as the main source for their stability and complex density structure. However numerical hydro- and
magneto-hydrodynamical simulations show that supersonic turbulence dissipates on timescales shorter
than the collapse timescale (Stone et al. 1998; Mac Low et al. 1998). In addition, no driver
of molecular cloud turbulence has ever been found which on the one hand suppresses
star formation on the small scales while, at the same time, stabilizing giant molecular clouds 
on the large scales (Heitsch et al. 2001). 

Another serious problem is the so called post-T Tauri problem (Hartmann 2001, 2002, 2003;
Hartmann et al. 2001): the typical age spread of young stellar populations is
of the order of 1-3 Myrs which is surprisingly narrow, indicating a coherent star 
formation process (see however Palla et al. 2005).  Figure 3, for example, shows
the young stars observed in the Taurus complex. They are aligned in 3 parallel filaments at the ridges
of an irregular diffuse gas complex. The origin of this alignment is not understood. In addition,
the age spread of the stars is only a few Myrs which is a factor of 10 smaller than the lateral
sound crossing times of the filaments. Which processes triggered star formation coherently along
all three filaments at exactly the same time? Finally, almost all molecular clouds in the solar 
neighborhood show signs of star formation which implies that they cannot be much older than
a few Myrs as otherwise either the newly formed stellar systems should have a larger age spread
or a larger fraction of clouds should not show signatures of star formation.

If molecular clouds condense into stars within a few Myrs the problem arises how
clouds with masses of $10^3 - 10^6$ M$_{\odot}$ could form in the first place. 
Consider a perturbation travelling through the ISM and sweeping up gas.
Adopting typical velocities of order 10 km/s, cross sections of order (10 pc)$^2$ and densities
of the diffuse gas  that is being swept up of order 1 cm$^{-3}$ it takes more than 
$10^7$ yrs to accumulate a total mass
of $10^3 - 10^4$  M$_{\odot}$. Giant molecular clouds might form in larger-scale
spiral arms with dimensions of several 100 pc and speeds of order 50 - 100 km/s
(Bonnell et al. 2006). Still, several $10^7$ yrs are required which is in conflict with the
above mentioned molecular cloud lifetimes.

\subsection{Formation of molecular clouds in converging gas flows}

The arguments, presented in the previous section, indicate that molecular cloud
formation and evolution is highly dynamical and that clouds never have the time
to achieve a long-term equilibrium state,
supported by internal turbulence. They instead form from the turbulent diffuse ISM with their
irregular motions and filamentary substructures probably already imprinted at the time of formation.
They are dispersed a few dynamical timescales later by star formation. That molecular clouds 
are transient dense islands in a turbulent sea and part of
a hierarchy of structures that form in the ubiquitous
colliding shock waves of the turbulent, diffuse ISM has already been outlined a 
long time ago by von Weizs\"acker (1951, for a summary of early work 
in this field see Elmegreen \& Scalo 2004). At that time however it was not possible
to study this model with numerical methods in details.

\begin{figure}
\resizebox{16cm}{!}{
{\includegraphics{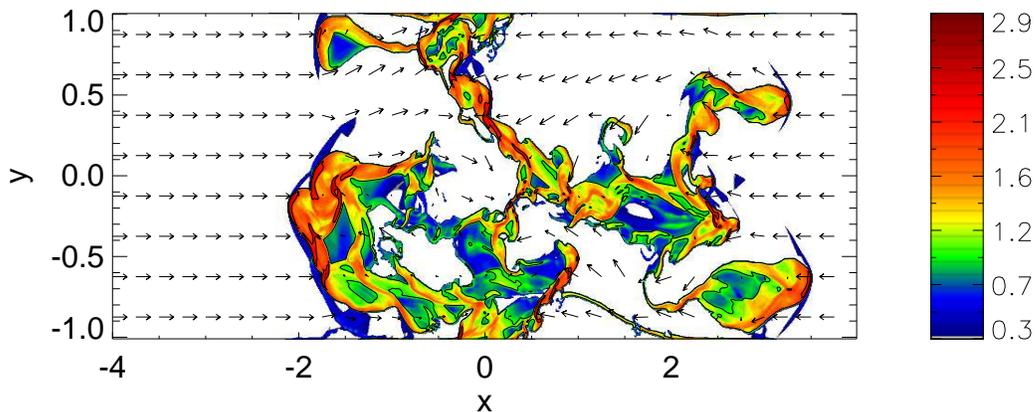}} 
}
\caption{Formation of a turbulent, irregular dense sheet of cold gas
in the interaction zone of two colliding gas flows with inflow
velocity of 10 km/s and temperature of 3000 K. The density distribution of
the cold gas with a temperature of $\leq 300$ K is shown. The colorbar
shows the logarithm of the particle density in units of
cm$^{-3}$. Vectors show the velocity of the diffuse intercloud medium that
moves into the inner region of the cold gas complex along open channels,
driving irregular turbulent flows in its interior.}
\label{fig4}
\end{figure}

The situation has changed in the meantime due to sophisticated numerical schemes
and the availability of fast computers.
The cooling and subfragmentation of dense slabs that form in colliding gas flows
has now been investigated in details
(Walder \& Folini 1998, 2000; Klein \& Woods 1998; 
Heitsch et al. 2005, 2006; V\'{a}zquez-Semadeni et al. 2005; Bonnell et al. 2006). 
It has been shown that the 
compressed region fragments as a result of a combination of thermal- and hydrodynamical instabilities 
(non-linear thin shell instability: Vishniac 1994; Kelvin-Helmholtz instability: Blondin \& Marks 1996:
thermal instability: Burkert \& Lin 2000; see also Hennebelle \& P\'{e}rault 1999, 2000;
Koyama \& Inutsuka 2004). Figure 4 shows an example
of a fragmenting, cold compressed slab. Dense filamentary and clumpy HI substructures
with temperatures of order 100 K
are forming that move with irregular velocities in a diffuse interclump medium. 
The numerical simulations show that even with modest
inflow speeds and completely uniform inflows, non-linear density perturbations form
that could represent the seeds of structure in molecular clouds.
These irregular sheets and filaments might lateron collapse and fragment into stars 
(e.g. Burkert \& Bodenheimer 1993; Klessen \& Burkert 2000; 
Burkert and Hartmann 2004). The random gas velocities within individual cold clumps are quite small.
However the relative velocities of the cold gas clumps with respect to each other 
are highly supersonic, compared to the sound speed of the cold component
and consistent with the observed values of a few km/s. The irregular,
turbulent motion of the cold gas in the slab is continuously driven by the kinetic energy 
of the inflowing diffuse gas. This might stabilize the region against 
gravitational collapse as long as the inflow continues. 

If the inflowing gas is preferentially HI, a dense irregular HI slab forms.
The formation of molecular gas
starts as soon as the HI surface density reaches values of $\Sigma_{crit} \approx 10^{21}$ cm$^{-2}$ 
(Bergin et al. 2004). 
For inflow speeds of 10 km/s this will require timescales of order a
few $10^7$ yrs for densities of the inflowing material of 1 cm$^{-3}$. The 
accumulated molecular cloud
mass would then be of order 1000 M$_{\odot}$ which is typical for filamentary molecular
clouds like Taurus in the solar neighborhood. Similar timescales are required in
the larger-scale flows in spiral density waves that produce giant molecular cloud
complexes. During the early period of proto-cloud evolution,
the gas would still be atomic and not easily detectable.  Bergin et al. (2004) show 
however that as soon as $\Sigma_{crit}$ is reached, the conversion into a molecular cloud
takes only a few $10^6$ yrs. In addition, the previously 100 K gas would cool down to 
10 K which could trigger local gravitational collapse and star formation. More work
is required to detect this gas phase observationally and investigate its evolution
with numerical simulations. 

\section{Summary}
The interstellar medium is a highly turbulent mixture of various interacting gas phases.
The existence of turbulence is expected, given the fact that
the typical Reynolds numbers in the ISM are of order $10^4 - 10^7$ (Elmegreen \& Scalo 2004).
That turbulence is however also found in regions without energy input by stars is 
surprising.

A wealth of connected and coupled structures on different scales all the way down to the 
scale of star formation have been identified, starting with the
early work of Larson (1981) who found that the density and velocity dispersion of 
molecular clouds scales with the size of the region as a power law. This indicates
an interesting connection between ISM dynamics and star formation. 
Many questions are still unsolved and need to be expained theoretically.
This might eventually lead to a quantitative model
of star formation that is one of the crucial missing ingredients in order to understand the
evolution of galaxies.

{\bf\ Acknowledgements:} I would like to acknowledge inspiring discussions with 
Lee Hartmann, Bruce Elmegreen, Fabian Heitsch, Nick Scoville and Sami Dib on this topic
and would like to thank Sungeun Kim, Lee Hartmann and Bruce Elmegreen 
for providing figures for this review.



\begin{thebibliography}{}


\bibitem[Alcock et al (2001)]{Alcock:2001} Alcock C. et al: 2001, ApJS 136, 439
\bibitem[Bahcall (1984)]{Bahcall:1984} Bahcall J.N.: 1984, ApJ 287, 926
\bibitem[Balbus et al (1991)]{Balbus:1991} Balbus, S.A., Hawley, J.F.: 1991, ApJ 376, 214
\bibitem[Ballesteros et al (2006)]{Ballesteros:2006} Ballesteros-Paredes, J., Klessen, R.S.,
Mac Low, M.-M., Vazquez-Semadeni, E.: 2006, to appear in Protostars and Planets V, astro-ph/0603357
\bibitem[Bergin et al (2004)]{Bergin:2004} Bergin, E.A., Hartmann, L.W., Raymond, J.C.,
Ballesteros-Paredes, J.: 2004, ApJ 612, 921
\bibitem[Blitz et al (1980)]{Blitz:1980} Blitz, L., Shu, F.H.: 1980, ApJ 238, 148
\bibitem[Blondin et al (1996)]{Blondin:1996} Blondin, J.M., Marks, B.S.: 1996, New Astronomy 1, 235
\bibitem[Bonnell et al (2006)]{Bonnell:2006} Bonnell, I.A., Dobbs, C.L., Robitaille, 
T.P., Pringle, J.E.: 2006, MNRAS 365, 37
\bibitem[Burkert (1993)]{Burkert:1993} Burkert, A., Bodenheimer, P.: 1993, MNRAS 264, 798
\bibitem[Burkert (2000)]{Burkert:2000} Burkert, A., Lin, D.N.C.: 2000, ApJ 537, 270
\bibitem[Burkert (2004)]{Burkert:2004} Burkert, A., Hartmann, L.: 2004, ApJ 616, 288
\bibitem[Cappellaro et al (1999)]{Cappellaro:1999} Cappellaro, E., Evans, R., Turatto, M.: 1999, A \& A 351, 459
\bibitem[Cowie et al (1995)]{Cowie:1995} Cowie, L., Hu, E., Songalia, A.: 1995, AJ 110, 1576
\bibitem[de Avillez et al (2004)]{Avillez:2004} de Avillez, M.A., Breitschwerdt, D.: 2004, A \& A 425, 899
\bibitem[de Avillez et al (2005)]{Avillez:2005} de Avillez, M.A., Breitschwerdt, D.: 2005, ApJ 634, L65
\bibitem[Dib et al (2005)]{Dib:2005} Dib, S., Burkert, A.: 2005, ApJ 630, 238
\bibitem[Dib et al (2006)]{Dib:2006} Dib, S., Bell, E., Burkert, A.: 2006, ApJ 638, 797
\bibitem[Dickey et al (1977)]{Dickey:1977} Dickey, J.M., Salpeter, E.E., Terzian, Y.: 1977, ApJ 211, L77
\bibitem[Dickey et al (1990)]{Dickey:1990} Dickey, J.M., Lockman, F.J.: 1990, ARAA 28, 215
\bibitem[Dziourkevitch et al (2004)]{Dziourkevitch:2004} Dziourkevitch, N., Elstner, D., R\"udiger, G.:
2004, A \& A 423, 29
\bibitem[Elmegreen (1989)]{Elmegreen:1989} Elmegreen, B.G.: 1989, ApJ 347, 859
\bibitem[Elmegreen (2001)]{Elmegreen:2001} Elmegreen, B.G., Kim, S., Staveley-Smith, L.: 2001, ApJ 548, 749
\bibitem[Elmegreen (2004)]{Elmegreen:2004} Elmegreen, B.G.: 2004, In Penetrating bars
through the masks of cosmic dust: the Hubble tuning fork strikes a new note, ed. D.L. Block,
I. Puevari, K.C. Freeman, R. Groess, E.K. Block, Astrophys. Space Sci. 319, 561
\bibitem[Elmegreen et al (2004a)]{Elmegreen:2004a} Elmegreen, B.G., Scalo, J.: 2004, ARAA 42, 211
\bibitem[Elmegreen et al (2004b)]{Elmegreen:2004b} Elmegreen, D.M., Elmegreen, B.G., Hirst, A.C.:
2004, ApJ 604, L21
\bibitem[Falgarone et al (1996)]{Falgarone:1996} Falgarone, E., Phillips, T.G.: 1996, ApJ 472, 191
\bibitem[Field (1965)]{Field:1965} Field, G.B.: 1965, ApJ 142, 531
\bibitem[Field et al (1969)]{Field:1969} Field, G.B., Goldsmith, D.W., Habing, H.J.: 1969, ApJ 155, 149
\bibitem[Forster Schreiber et al (2006)]{Forster:2006} Forster Schreiber, N.M. et al: 2006, ApJ in press,
astro-ph/0603559
\bibitem[Gazol et al (2001)]{Gazol:2001} Gazol, A., V\'{a}zquez-Semadeni, E., S\'{a}nchez-Salcedo, F.J.,
Scalo, J.: 2001, ApJ 557, L121
\bibitem[Hartmann (2001)]{Hartmann:2001a} Hartmann, L.: 2001, AJ 121, 1030
\bibitem[Hartmann (2002)]{Hartmann:2002} Hartmann, L.: 2002, ApJ 578, 914
\bibitem[Hartmann (2003)]{Hartmann:2003} Hartmann, L.: 2003, ApJ 585, 398
\bibitem[Hartmann et al (2001)]{Hartmann:2001b} Hartmann, L., Ballesteros-Paredes, J.,
Bergin, E.A.: 2001, ApJ 562, 852
\bibitem[Heiles (2001)]{Heiles:2001} Heiles, C.: 2001, ApJ 551, L105
\bibitem[Heitsch (2001)]{Heitsch:2001} Heitsch, F., Mac Low, M., Klessen, R.S.: 2001, ApJ 547, 280
\bibitem[Heitsch (2005)]{Heitsch:2005} Heitsch, F., Burkert, A., Hartmann, L., Slyz, A.D., Devriendt, J.E.:
2005, ApJ 633, L113
\bibitem[Heitsch (2006)]{Heitsch:2006} Heitsch, F., Slyz, A.D., Devriendt, J.E.G., Hartmann, L.,
Burkert, A.: 2006, ApJ, submitted (astro-ph/????)
\bibitem[Hennebelle et al (1999)]{Hennebelle:1999} Hennebelle, P., P\'{e}rault, M.: 1999, A \& A 351, 309
\bibitem[Hennebelle et al (2000)]{Hennebelle:2000} Hennebelle, P., P\'{e}rault, M.: 2000, A \& A 359, 1124
\bibitem[Jenkins (2004)]{Jenkins:2004} Jenkins, J.B.: 2004, Astrophys. Space Sci. 289, 215
\bibitem[Klein (1998)]{Klein:1998} Klein, R.I., Woods, D.T.: 1998, ApJ 497, 777
\bibitem[Kanekar et al(2003)]{Kanekar:2003} Kanekar, N. C., Subrahmanyan, R., Chengular, J.N.,
Safouris, V.: 2003, MNRAS 346, L57
\bibitem[Kerp et al (2002)]{Kerp:2002} Kerp, J., Walter, F., Brinks, E.: 2002, ApJ 571, 809
\bibitem[Kessel-Deynet et al (2003)]{Kessel:2003} Kessel-Deynet, O., Burkert, A.: 2003, MNRAS 338, 545
\bibitem[Kim et al (1999)]{Kim:1999} Kim, S., Dopita, M.A., Staveley-Smith, L., Bessell, M.S.: 1999,
AJ 118, 2797
\bibitem[Kim et al (2001)]{Kim:2001} Kim, J., Balsara, D., Mac Low, M.M.: 2001, JKAS 34, 333
\bibitem[Kim et al (2003)]{Kim:2003} Kim, W.T., Ostriker, E., Stone, J.: 2003, ApJ 599, 1157
\bibitem[Klessen et al (2000)]{Klessen:2000} Klessen, R.S., Burkert, A.: 2000, ApJS 128, 287
\bibitem[Kolmogorov (1941)]{Kolmogorov:1941} Kolmogorov, A.N.: 1941, Proc. R. Soc. London Ser. A 434, 9
\bibitem[Koyama et al (2004)]{Koyama:2004} Koyama, H., Inutsuka, S.: 2004, RMxAC 22,26
\bibitem[Kritsuk et al (2002b)]{Kritsuk:2002b} Kritsuk, A.G., Norman, M.L.: 2002b, ApJ 569, L127
\bibitem[Kritsuk et al (2002a)]{Kritsuk:2002a} Kritsuk, A.G., Norman, M.L.: 2002a, ApJ 580, L51
\bibitem[Larson (1981)]{Larson:1981} Larson, R.B.: 1981, MNRAS 194, 809
\bibitem[MacLow et al (1998)]{MacLow:1998} Mac Low, M.M., Klessen, R.S., Burkert, A.,
Smith, M.D.: 1998, Phys. Rev. Lett. 80, 2754
\bibitem[MacLow et al (2004)]{MacLow:2004} Mac Low, M.M., Klessen, R.S.: 2004, Rev. Mod. Phys. 76, 125
\bibitem[MacLow et al (2005)]{MacLow:2005} Mac Low, M.M., Balsara, D.S., Kim, J., de Avillez, M.A.: 
2005, ApJ 626, 864
\bibitem[McKee et al (1977)]{McKee:1977} McKee, C.F., Ostriker, J.P.: 1977, ApJ 218, 148
\bibitem[Palla (2005)]{Palla:2005} Palla, F., Randich, S., Flaccomio, E., Pallavicini, R.:
2005, ApJ 626, L49
\bibitem[Piontek et al (2004)]{Piontek:2004} Piontek, R., Ostriker, E.C.: 2004, ApJ 601, 905
\bibitem[Puche et al (1992)]{Puche:1992} Puche, D., Westpfahl, D.J., Brinks, E., Roy, J.R.: 
1992, AJ 103, 1841
\bibitem[Rhode et al (1999)]{Rhode:1999} Rhode, K.L., Salzer, J.J., Westpfahl, D.J., Radice, L.A.: 1999, 
AJ 118, 323
\bibitem[Scalo et al (2004)]{Scalo:2004} Scalo, J., Elmegreen, B.G.: 2004, ARAA 42, 275
\bibitem[Scalo (1987)]{Scalo:1987} Scalo, J.: 1987, In Interstellar Processes, ed. D.J.
Hollenbach, H.A. Thronson Jr, (Dordrecht: Reidel), p. 349
\bibitem[Sellwood et al (1999)]{Sellwood:1999} Sellwood, J.A., Balbus, S.A.: 1999, ApJ 511, 660
\bibitem[Slyz et al (2005)]{Slyz:2005} Slyz, A.D., Devriendt, J.E., Bryan, G., Silk, J.: 2005, MNRAS 356, 737
\bibitem[Stone et al (1998)]{Stone:1998} Stone, J.M., Ostriker, E.C., Gammie, C.F.: 1998, ApJ 508, L99
\bibitem[Tran et al (2003)]{Tran:2003} Tran, H. et al.: 2003, ApJ 585, 750
\bibitem[Vazquez (2000)]{Vazquez:2000} V\'{a}zquez-Semadeni, E., Ostriker, E.C., Passot, T.,
Gammie, C.F., Stone, J.M.: 2000, In Protostars and Planets IV, ed. V. Mannings, A.P.
Boss, S.S. Russell, (Tucson: Univ. Arizona), p.3
\bibitem[Vazquez-Semadeni et al (2005)]{Vazquez:2005} V\'{a}zquez-Semadeni, E., Ryu, D., Passot, T.,
G\'{o}nzalez, R.F., Gazol, A.: 2005, submitted
\bibitem[Vishniac (1994)]{Vishniac:1994} Vishniac, E.T.: 1994, ApJ 428, 186
\bibitem[Weizs\"acker (1951)]{Weizsaecker:1951} von Weizs\"acker, C.F.: 1951, ApJ 114, 165
\bibitem[Wada et al (2002)]{Wada:2002} Wada, K., Meurer, G., Norman, C.A.: 2002, ApJ 577, 197
\bibitem[Walder et al (1998)]{Walder:1998} Walder, R., Folini, D.: 1998, ApSS 260, 215
\bibitem[Walder et al (2000)]{Walder:2000} Walder, R., Folini, D.: 2000, ApSS 274, 343
\bibitem[Williams et al (2000)]{Williams:2000} Williams, J.P., Blitz, L., McKee, C.F.: 2000, In
Protostars and Planets IV, ed. V. Mannings , A.P.  Boss, S.S. Russell, (Tucson: Univ. Arizona), p.97

\end{thebibliography}
\end{document}